\documentclass[11pt]{article}

\begin{document}

\noindent
{\bf History of the BCS to Bose-Einstein Transition}

\noindent
Although I am partially rectifying the situation by sending photocopies
of a 1969 paper of mine$^{1}$ to at least one author of papers I notice
which state incorrectly or imply that the crossover between BCS 
and Bose-Einstein condensation theory was first studied by Leggett in
1980, there is still no general awareness of my work on this topic
published eleven years earlier.  An example of a paper which implies
incorrect history is that by Kleppner in a recent "Physics
Today".$^{2}$

\noindent
In the mid 1980's, colleagues of mine in Australia and myself also
found the first example of a system which lay on the Bose-gas side of
the BCS-Bose gas transition, in a ceramic sample of 3\% Zr-doped
SrTiO$_3$,$^{3-5}$ although with a pairing temperature much lower than
that predicted for the model used for this material in 1969.  In three
dimensions there is a threshold in the coupling strength for the
existence of Bose-gas superconductivity at very low carrier
concentrations, and, for strengths slightly above the threshold, the
pair binding energy is proportional to the square of the difference of
the coupling from the threshold value.  Thus, for example, an error in
calculated coupling giving a value 5\% above threshold instead of 1\%
above threshold would make a factor of 25 error in the pair binding
energy.

\noindent
An attempt was made later$^{6}$ to reach the Bose-gas r\'{e}gime in
macrocrystalline samples of Zr-doped SrTiO$_3$ with similar carrier
concentrations ($\sim 10^{15}$cm$^{-3}$ at low temperatures) to
those in the ceramic sample mentioned, but, unfortunately, in this case
some non-uniform state formed, and prevented the Bose-gas r\'{e}gime
being reached.  However, there is still much interesting physics
awaiting study in this material.

\noindent
{\bf References}

\noindent
1. D.M. Eagles, {\em Phys. Rev.} {\bf 186}, 456 (1969).\newline
2. D. Kleppner, {\em Physics Today}, August 2004, p. 12.\newline
3. R.J. Tainsh, C. Andrikidis, {\em Solid State Commun.} {\bf 60},
517 (1986).\newline
4. D.M. Eagles, {\em Solid State Commun.} {\bf 60}, 521 (1986). 
\newline     
5. D.M. Eagles, R.J. Tainsh, C. Andrikidis,  {\em Physica C} {\bf 157}, 48
(1989).\newline
6. C. Andrikidis, R.J. Tainsh, D.M. Eagles, {\em Physica B} {\bf 165-166}, 
1517 (1990).
       
{\bf D.M. Eagles} 

(e-mail: d.eagles@ic.ac.uk)

\noindent
{\em 19, Holt Road, Harold Hill, Romford, Essex RM3 8PN, England.}
\end{document}